# Is the Regge Trajectory Quasi-linear or Square-root Form?

Zhen Li [1], and Ke-Wei Wei [2, 3]

[1] *College of Physical Science and Technology*, *Yili Normal University*, *Yining 835000*, *China*

[2] *College of Physics and Electrical Engineering*, *Anyang Normal University*, *Anyang 455000*, *China*

[3] *Kavli Institute for Theoretical Physics China*, *CAS*, *Beijing 100190*, *China*

Correspondence should be addressed to Ke-Wei Wei; kewei_wei@163.com

**Abstract**

There are many orbital excited mesons discovered in recent years. In this work we attempt to study whether the Regge trajectory is quasi-linear or square-root form. In the framework of the quasi-linear Regge trajectory and square-root Regge trajectory, the masses of the states lying on the well established $1^1S_0$, $1^3S_1$, and $1^3P_2$ trajectories are estimated. Comparison of the results given by the two trajectories with the existing experimental data illustrates that both of them can give a reasonable description for the ground mesons. For the orbital excited states, the quasi-linear trajectory describes the existing meson spectrum to be more reasonable.

## 1. Introduction

The Regge theory descends from the analysis of the scattering amplitude properties in the complex orbital momentum plane [1], which is concerned with the particle spectrum, the forces between particles and the high energy behavior of scattering amplitudes [2]. Currently, the Regge trajectory is a simple and effective phenomenological model to study the hadron spectrum. A series of papers [3-14] show that meson states fit to quasi-linear Regge trajectories with sufficiently good accuracy. However, in Refs. [15-20], it was suggested that the realistic Regge trajectories could be non-linear. Among them, a most typical non-linear form-square-root trajectory was proposed in Ref. [20]. Then, it is a puzzle that the Regge trajectory is quasi-linear or square-root form.

In the recent issue of Review of Particle Physics [21], there are many orbital excited mesons have been established well, e.g., the orbital excited states $\rho_5(2350)$ and $K_5^*(2380)$ are assigned to $n\bar{n}$ and $n\bar{s}$ of $1^3G_5(5^{--})$, $a_6(2450)$ is assigned to $n\bar{n}$ of $1^3H_6(6^{++})$, and $D_{s3}^*(2860)$ is assigned to $c\bar{s}$ of $1^3D_3(3^{--})$. In the presence of these assignments, according to the newest experimental data from the Particle Data Group (PDG) [21], we respectively calculate the masses of orbital excitations by the quasi-linear trajectory and square-root trajectory. Comparing the results given by the quasi-linear trajectory and square-root trajectory with the experimental data, we examine whether Regge trajectory is quasi-linear or square-root form.

In the present work, we only calculate the mass spectrum of the $1^1S_0$, $1^3S_1$, and $1^3P_2$ trajectories, since these trajectories are of the rich available experimental data. The difference between the total spins of hadrons lying on the same Regge trajectory is $2n$ ($n$ = 1, 2, 3, …). Mesons with the quantum numbers $N^{2S+1}L_J$, $N^{2S+1}(L+2)_{J+2}$, $N^{2S+1}(L+4)_{J+4}$, $N^{2S+1}(L+6)_{J+6}$, … (where $N$, $L$ and $S$ denote the radial excited quantum number, the orbital excited quantum number and the intrinsic spin, respectively) lying on the same Regge trajectory [1, 5, 6, 22].

## 2. The Quasi-linear Regge Trajectory

By assuming the existence of the quasi-linear trajectories for a meson multiplet, one can have

$$J = \alpha_{i\bar{i'}}(0) + \alpha'_{i\bar{i'}} M_{i\bar{i'}}^2, \qquad (1)$$

where $i$ ($\bar{i'}$) refers to the quark (anti-quark) flavor, $J$ and $M_{i\bar{i'}}$ are respectively the spin and mass of the $i\bar{i'}$ meson, $\alpha_{i\bar{i'}}(0)$ and $\alpha'_{i\bar{i'}}$ are respectively the intercept and slope of the trajectory on which the $i\bar{i'}$ meson lies. The intercept and slope parameters for different flavors can be related by the following relations:

1) additivity of intercepts [5, 23-26],

$$\alpha_{i\bar{i}}(0) + \alpha_{j\bar{j}}(0) = 2\alpha_{i\bar{j}}(0), \qquad (2)$$

2) additivity of inverse slopes [5, 23],

$$\frac{1}{\alpha'_{i\bar{i}}} + \frac{1}{\alpha'_{j\bar{j}}} = \frac{2}{\alpha'_{i\bar{j}}}. \qquad (3)$$

These two additivity requirements are independent of which specific form is assumed for the trajectories [20].

## 3. The Square-root Regge Trajectory

In the analysis of Ref. [20], the specific square-root Regge trajectory which contacts with the spin $J$ and mass $M$ of a meson has been proposed as follows:

$$J = \beta_{i\bar{j}}(0) + \gamma(\sqrt{T_{i\bar{j}}} - \sqrt{T_{i\bar{j}} - M_{i\bar{j}}^2}), \tag{4}$$

where $i(\bar{j})$ refers to the quark (anti-quark) flavor, $J$ and $M_{i\bar{j}}$ are respectively the spin and mass of the $i\bar{j}$ meson, $\gamma$ is a constant independent of meson flavor. $\beta_{i\bar{j}}(0)$ is the intercept of the Regge trajectory. When $M_{i\bar{j}}$ reaches $\sqrt{T_{i\bar{j}}}$, the real part of the square-root trajectory stops growing, and there are no states with a higher spin quantum number than $J_{\max} = [J(T_{i\bar{j}})]$. The parameter $T_{i\bar{j}}$ is therefore the trajectories termination point and $\sqrt{T_{i\bar{j}}}$ is the threshold parameter of meson trajectories.

Note that for $M \ll T$, according to the Taylor series expansion, Eq. (4) reduces to the quasi-linear form

$$J \approx \beta_{i\bar{j}}(0) + \frac{\gamma}{2\sqrt{T_{i\bar{j}}}} M_{i\bar{j}}^2 = \beta_{i\bar{j}}(0) + \beta'_{i\bar{j}}(0) M_{i\bar{j}}^2. \tag{5}$$

For a meson multiplet, the parameters for different flavors can be related by the following relations:

1) additivity of intercepts,

$$\beta_{i\bar{i}}(0) + \beta_{j\bar{j}}(0) = 2\beta_{i\bar{j}}(0). \tag{6}$$

2) On the basis of additivity of inverse slopes near the origin of the square-root trajectory, with the help of the relation (5), one can get the following relation

$$\sqrt{T_{i\bar{i}}} + \sqrt{T_{j\bar{j}}} = 2\sqrt{T_{i\bar{j}}}. \tag{7}$$

## 4. The Meson Spectrum

4.1 The Meson Spectrum of the $1^3S_1$ Trajectories

According to PDG, the states $\rho$, $K^*(892)$, $D^*$, $J/\psi$, $B^*$ and $\Upsilon$ [1] belong to the members of the $1^3S_1$ meson multiplet, $\rho_3(1690)$ and $K_3^*(1780)$ belong to the $1^3D_3$ meson multiplet, inserting the masses of these mesons into the following equations

---

[1] $M_{K^*(892)} = (M_{K^*(892)^\pm} + M_{K^*(892)^0})/2$, $M_{D^*} = (M_{D^*(2010)^\pm} + M_{D^*(2007)^0})/2$. Here and below, all masses of the well-established states used as input are taken from PDG [21].

$$1 = \alpha_{n\bar{n}}(0) + \alpha'_{n\bar{n}} M_\rho^2, \tag{8}$$

$$3 = \alpha_{n\bar{n}}(0) + \alpha'_{n\bar{n}} M_{\rho_3}^2, \tag{9}$$

$$1 = \alpha_{n\bar{s}}(0) + \alpha'_{n\bar{s}} M_{K^*(892)}^2, \tag{10}$$

$$3 = \alpha_{n\bar{s}}(0) + \alpha'_{n\bar{s}} M_{K_3^*(1780)}^2, \tag{11}$$

$$1 = \alpha_{c\bar{c}}(0) + \alpha'_{c\bar{c}} M_{J/\psi}^2, \tag{12}$$

$$1 = \alpha_{c\bar{n}}(0) + \alpha'_{c\bar{n}} M_{D^*}^2, \tag{13}$$

$$1 = \alpha_{b\bar{b}}(0) + \alpha'_{b\bar{b}} M_\Upsilon^2, \tag{14}$$

$$1 = \alpha_{n\bar{b}}(0) + \alpha'_{n\bar{b}} M_{B^*}^2, \tag{15}$$

with the help of the relations (2) and (3), one can reckon the Regge intercept and Regge slope of the $1^3S_1$ trajectories. These parameters are summarized in Table 1.

TABLE 1: Parameters of the $1^3S_1$ trajectories of the form (1).

|  | $n\bar{n}$ | $n\bar{s}$ | $s\bar{s}$ | $c\bar{c}$ | $c\bar{n}$ |
|---|---|---|---|---|---|
| $\alpha(0)$ | 0.4660 | 0.3218 | 0.1776 | -3.1998 | -1.3669 |
| $\alpha'/\text{GeV}^{-2}$ | 0.8885 | 0.8491 | 0.8130 | 0.4379 | 0.5867 |
|  | $c\bar{s}$ | $b\bar{b}$ | $n\bar{b}$ | $s\bar{b}$ | $c\bar{b}$ |
| $\alpha(0)$ | -1.5110 | -17.4107 | -8.4724 | -8.6166 | -10.3052 |
| $\alpha'/\text{GeV}^{-2}$ | 0.5692 | 0.2057 | 0.3341 | 0.3283 | 0.2799 |

Using the parameters shown in Table 1, with the help of the relation (1), we calculate masses of the spin-1, spin-3, spin-5, and spin-7 mesons lying on these trajectories. The results obtained by the quasi-linear trajectory are shown in Table 3. In Table 3 and subsequent Table 6 and Table 9, the values used as input for our analysis are shown in boldface. The results extracted by the quasi-linear trajectory and square-root trajectory are compared with the experimental data in Table 3.

In the framework of the square-root trajectory, we start with the $\rho$ trajectory. The intercept $\beta_\rho(0)$ of this trajectory is well established. The intercept value was taken to be 0.55 in Ref. [20], which is consistent with the values extracted from the behavior of the differential cross section of the process $\pi^- p \to \pi^0 n$ [27-29] and inferred from the difference of the total cross sections of $\pi^+ p$ and $\pi^- p$ scattering [30-31]. In addition, the intercept $\beta_\rho(0)$ was extracted to be 0.55 from the analysis of $pp$ and $\bar{p}p$ scattering data in a simple pole exchange model [32]. So we take the value of the intercept of $\rho$ trajectory to be 0.55. According to PDG, $\rho(770)$ and $\rho_3(1690)$ belong to the states lying on the $\rho$ trajectory. Resorting to the relation (4), one can have

$$1 = \beta_\rho(0) + \gamma(\sqrt{T_\rho} - \sqrt{T_\rho - M_\rho^2}), \tag{16}$$

$$3 = \beta_\rho(0) + \gamma(\sqrt{T_\rho} - \sqrt{T_\rho - M_{\rho_3}^2}). \tag{17}$$

Inserting the masses of $\rho$ and $\rho_3$ into the relations (16) and (17), one can extract the parameters $\gamma$ and $\sqrt{T_\rho}$:

$$\gamma = 3.4148 \text{ GeV}^{-1}, \quad \sqrt{T_\rho} = 2.3463 \text{ GeV}. \tag{18}$$

Resorting to the value of the constant $\gamma$ in Eq. (18), inserting the masses of $K^*(892)$, $D^*$, $J/\psi$, $B^*$, $\Upsilon$, and $K_3^*(1780)$ into the following equations

$$1 = \beta_{n\bar{s}}(0) + \gamma(\sqrt{T_{n\bar{s}}} - \sqrt{T_{n\bar{s}} - M_{K^*(892)}^2}), \tag{19}$$

$$3 = \beta_{n\bar{s}}(0) + \gamma(\sqrt{T_{n\bar{s}}} - \sqrt{T_{n\bar{s}} - M_{K_3^*(1780)}^2}), \tag{20}$$

$$1 = \beta_{c\bar{c}}(0) + \gamma(\sqrt{T_{c\bar{c}}} - \sqrt{T_{c\bar{c}} - M_{J/\psi}^2}), \tag{21}$$

$$1 = \beta_{c\bar{n}}(0) + \gamma(\sqrt{T_{c\bar{n}}} - \sqrt{T_{c\bar{n}} - M_{D^*}^2}), \tag{22}$$

$$1 = \beta_{b\bar{n}}(0) + \gamma(\sqrt{T_{b\bar{n}}} - \sqrt{T_{b\bar{n}} - M_{B^*}^2}), \tag{23}$$

$$1 = \beta_{b\bar{b}}(0) + \gamma(\sqrt{T_{b\bar{b}}} - \sqrt{T_{b\bar{b}} - M_\Upsilon^2}), \tag{24}$$

and by means of the relations (6) and (7), one can extract the intercept and threshold parameters of the $1^3S_1$ trajectories, these parameters are summarized in Table 2.

TABLE 2: Parameters of $1^3S_1$ trajectories of the form (4).

|                    | $n\bar{n}$ | $n\bar{s}$ | $s\bar{s}$ | $c\bar{c}$ | $c\bar{n}$ |
|---|---|---|---|---|---|
| $\beta(0)$         | **0.55**   | 0.4287     | 0.3074     | -2.5391    | -0.9945    |
| $\sqrt{T}$ / GeV   | 2.3463     | 2.4710     | 2.5957     | 5.1452     | 3.7458     |
|                    | $c\bar{s}$ | $b\bar{b}$ | $n\bar{b}$ | $s\bar{b}$ | $c\bar{b}$ |
| $\beta(0)$         | -1.1159    | -14.5144   | -6.9822    | -7.1035    | -8.5268    |
| $\sqrt{T}$ / GeV   | 3.8705     | 12.1211    | 7.2337     | 7.3584     | 8.6332     |

Using the parameters shown in Table 2, resorting to the relation (4), we can obtain masses of the spin-1, spin-3, spin-5, and spin-7 states lying on these trajectories. The results extracted by the square-root trajectory are also shown in Table 3.

TABLE 3: Comparison of the masses of the $J = 1, 3, 5, 7$ mesons lying on the $1^3S_1$ trajectories with the experimental data. (All in MeV.) The numbers in boldface are the experimental values taken as the input.

|  | $J = 1$ | | | $J = 3$ | | | $J = 5$ | | | $J = 7$ | | |
|---|---|---|---|---|---|---|---|---|---|---|---|---|
|  | Quasi-linear | Square-root | Exp.[21] | Quasi-linear | Square-root | Exp.[21] | Quasi-linear | Square-root | Exp.[21] | Quasi-linear | Square-root | Exp.[21] |
| $M_{n\bar{n}}$ | **775.26** | **775.26** | 775.26 | **1688.8** | **1688.8** | 1688.8 | 2259 | 2102 | 2330 | 2712 | 2301 | 2747 |
| $M_{n\bar{s}}$ | **893.74** | **893.74** | 893.74 | **1776** | **1776** | 1776 | 2347 | 2196 | 2382 | 2804 | 2410 | |
| $M_{c\bar{n}}$ | **2008.62** | **2008.62** | 2008.62 | 2728 | 2719 | | 3294 | 3173 | | 3776 | 3472 | |
| $M_{c\bar{s}}$ | 2100 | 2101 | 2112.1 | 2815 | 2807 | 2863.2 | 3382 | 3264 | | 3867 | 3571 | |
| $M_{c\bar{c}}$ | **3096.916** | **3096.916** | 3096.916 | 3763 | 3750 | | 4327 | 4224 | | 4826 | 4576 | |
| $M_{b\bar{n}}$ | **5324.83** | **5324.83** | 5324.83 | 5860 | 5809 | | 6350 | 6201 | | 6805 | 6517 | |
| $M_{b\bar{s}}$ | 5412 | 5412 | 5415.4 | 5949 | 5898 | | 6440 | 6293 | | 6897 | 6612 | |
| $M_{b\bar{c}}$ | 6355 | 6355 | | 6895 | 6848 | | 7395 | 7260 | | 7863 | 7605 | |
| $M_{b\bar{b}}$ | **9460.30** | **9460.30** | 9460.30 | 9961 | 9901 | | 10438 | 10290 | | 10894 | 10632 | |

4.2 The Meson Spectrum of the $1^1S_0$ Trajectories

The states $\pi$, $K$, $\eta_c(1S)$, $D$, $\eta_b(1S)$ and $B$ [2] belong to the members of the $1^1S_0$ meson multiplet, $\pi_2(1670)$ and $K_2(1770)$ belong to the $1^1D_2$ meson multiplet [21], inserting the masses and spins of these mesons into the relation (1), with the help of the relations (2) and (3), one can extract the intercept and slope parameters of $1^1S_0$ trajectories by the quasi-linear trajectory. The parameters are shown in Table 4.

TABLE 4: Parameters of $1^1S_0$ trajectories of the form (1).

|  | $n\bar{n}$ | $n\bar{s}$ | $s\bar{s}$ | $c\bar{c}$ | $c\bar{n}$ |
|---|---|---|---|---|---|
| $\alpha(0)$ | -0.01357 | -0.1695 | -0.3254 | -3.5947 | -1.8041 |
| $\alpha' / \text{GeV}^{-2}$ | 0.7201 | 0.6902 | 0.6627 | 0.4038 | 0.5174 |
|  | $c\bar{s}$ | $b\bar{b}$ | $n\bar{b}$ | $s\bar{b}$ | $c\bar{b}$ |
| $\alpha(0)$ | -1.9601 | -16.6177 | -8.3157 | -8.4716 | -10.1062 |
| $\alpha' / \text{GeV}^{-2}$ | 0.5018 | 0.1881 | 0.2983 | 0.2930 | 0.2566 |

---

[2] $M_\pi = (M_{\pi^\pm} + M_{\pi^0})/2$, $M_K = (M_{K^\pm} + M_{K^0})/2$, $M_D = (M_{D^\pm} + M_{D^0})/2$, $M_B = (M_{B^\pm} + M_{B^0})/2$

Using the parameters shown in Table 4, we calculate masses of the spin-0, spin-2, spin-4, and spin-6 mesons lying on these trajectories. The results extracted by the quasi-linear trajectory are shown in Table 6.

Inserting the masses and spins of $\pi$, $K$, $\eta_c(1S)$, $D$, $\eta_b(1S)$, $B$, $\pi_2(1670)$, and $K_2(1770)$ into the relation (4), resorting to the relations (5) and (6), one can extract the intercept and threshold parameters of $1^1S_0$ trajectories by the square-root trajectory. The parameters are summarized in Table 5. As demonstrated in Ref. [20], here, and subsequent $1^3P_2$ trajectories, the constant $\gamma$ is the same as $1^3S_1$ trajectories, since the constant $\gamma$ is independent of meson flavor.

TABLE 5: Parameters of $1^1S_0$ trajectories of the form (4).

|   | $n\bar{n}$ | $n\bar{s}$ | $s\bar{s}$ | $c\bar{c}$ | $c\bar{n}$ |
|---|---|---|---|---|---|
| $\beta(0)$ | -0.01207 | -0.1504 | -0.2887 | -3.1973 | -1.6047 |
| $\sqrt{T}$ / GeV | 2.6675 | 2.8108 | 2.9541 | 5.2219 | 3.9447 |
|   | $c\bar{s}$ | $b\bar{b}$ | $n\bar{b}$ | $s\bar{b}$ | $c\bar{b}$ |
| $\beta(0)$ | -1.7430 | -14.7808 | -7.3964 | -7.5348 | -8.9891 |
| $\sqrt{T}$ / GeV | 4.0880 | 12.3668 | 7.5171 | 7.6605 | 8.7944 |

Using the parameters shown in Table 5, resorting to the relation (4), one can obtain masses of the spin-0, spin-2, spin-4, and spin-6 states lying on these trajectories. The results given by the square-root trajectory are shown in Table 6.

TABLE 6: Comparison of the masses of the $J = 0, 2, 4, 6$ mesons lying on the $1^1S_0$ trajectories with the experimental data. (All in MeV.)

|   | $J = 0$ | | | $J = 2$ | | | $J = 4$ | | | $J = 6$ | | |
|---|---|---|---|---|---|---|---|---|---|---|---|---|
|   | Quasi-linear | Square-root | Exp.[21] | Quasi-linear | Square-root | Exp.[21] | Quasi-linear | Square-root | Exp.[21] | Quasi-linear | Square-root | Exp.[21] |
| $M_{n\bar{n}}$ | **137.273** | **137.273** | 137.273 | **1672.2** | **1672.2** | 1672.2 | 2361 | 2211 |   | 2890 | 2509 |   |
| $M_{n\bar{s}}$ | **495.644** | **495.644** | 495.644 | **1773** | **1773** | 1773 | 2458 | 2314 |   | 2990 | 2623 |   |
| $M_{c\bar{n}}$ | **1867.23** | **1867.23** | 1867.23 | 2712 | 2686 |   | 3349 | 3202 |   | 3884 | 3551 |   |
| $M_{c\bar{s}}$ | 1976 | 1978 | 1968.30 | 2809 | 2786 |   | 3446 | 3305 |   | 3983 | 3660 |   |
| $M_{c\bar{c}}$ | **2983.6** | **2983.6** | 2983.6 | 3722 | 3685 |   | 4337 | 4192 |   | 4875 | 4569 |   |
| $M_{b\bar{n}}$ | **5279.45** | **5279.45** | 5279.45 | 5881 | 5814 |   | 6425 | 6248 |   | 6928 | 6602 |   |
| $M_{b\bar{s}}$ | 5377 | 5379 | 5366.79 | 5978 | 5915 |   | 6524 | 6352 |   | 7028 | 6709 |   |
| $M_{b\bar{c}}$ | 6276 | 6275 | 6275.1 | 6869 | 6800 |   | 7414 | 7241 |   | 7923 | 7612 |   |
| $M_{b\bar{b}}$ | **9398.0** | **9398.0** | 9398.0 | 9949 | 9869 |   | 10470 | 10285 |   | 10966 | 10653 |   |

## 4.3 The Meson Spectrum of the $1^3P_2$ Trajectories

The states $a_2(1320)$, $K_2^*(1430)$, $\chi_{c2}(1P)$, $D_2^*(2460)$, $\chi_{b2}(1P)$ and $B_2^*(5747)$ [3] belong to the members of the $1^3P_2$ meson multiplet, $a_4(2040)$ and $K_4^*(2045)$ belong to the $1^3F_4$ meson multiplet [21], inserting the masses and spins of these mesons into the relation (1), together with the relations (2) and (3), one can extract the intercept and slope parameters of $1^3P_2$ meson trajectories by the quasi-linear trajectory. The parameters are shown in Table 7.

TABLE 7: Parameters of $1^3P_2$ trajectories of the form (1).

|  | $n\bar{n}$ | $n\bar{s}$ | $s\bar{s}$ | $c\bar{c}$ | $c\bar{n}$ |
|---|---|---|---|---|---|
| $\alpha(0)$ | 0.4498 | 0.09154 | -0.2667 | -3.8158 | -1.6830 |
| $\alpha'$ / GeV$^{-2}$ | 0.8920 | 0.9346 | 0.9815 | 0.4599 | 0.6069 |
|  | $c\bar{s}$ | $b\bar{b}$ | $n\bar{b}$ | $s\bar{b}$ | $c\bar{b}$ |
| $\alpha(0)$ | -2.0413 | -20.1508 | -9.8505 | -10.2088 | -11.9833 |
| $\alpha'$ / GeV$^{-2}$ | 0.6263 | 0.2254 | 0.3599 | 0.3666 | 0.3025 |

Using the parameters shown in Table 7, we calculate masses of the spin-2, spin-4, spin-6, and spin-8 mesons lying on these trajectories. The results given by the quasi-linear trajectory are shown in Table 9.

Inserting the masses and spins of $a_2(1320)$, $K_2^*(1430)$, $\chi_{c2}(1P)$, $D_2^*(2460)$, $\chi_{b2}(1P)$, $B_2^*(5747)$, $a_4(2040)$ and $K_4^*(2045)$ into the relation (4), resorting to the relations (5) and (6), one can extract the intercept and threshold parameters of $1^3P_2$ trajectories by the square-root trajectory. The parameters are summarized in Table 8.

TABLE 8: Parameters of $1^3P_2$ trajectories of the form (4).

|  | $n\bar{n}$ | $n\bar{s}$ | $s\bar{s}$ | $c\bar{c}$ | $c\bar{n}$ |
|---|---|---|---|---|---|
| $\beta(0)$ | 0.7578 | 0.5088 | 0.2598 | -2.6598 | -0.9510 |
| $\sqrt{T}$ / GeV | 2.5707 | 2.5564 | 2.5421 | 5.3161 | 3.9434 |
|  | $c\bar{s}$ | $b\bar{b}$ | $n\bar{b}$ | $s\bar{b}$ | $c\bar{b}$ |
| $\beta(0)$ | -1.2000 | -15.7488 | -7.4955 | -7.7445 | -9.2043 |
| $\sqrt{T}$ / GeV | 3.9291 | 12.0504 | 7.3106 | 7.2963 | 8.6833 |

---

[3] $M_\pi = (M_{\pi^\pm} + M_{\pi^0})/2$, $M_K = (M_{K^\pm} + M_{K^0})/2$, $M_D = (M_{D^\pm} + M_{D^0})/2$, $M_B = (M_{B^\pm} + M_{B^0})/2$.

Using the parameters shown in Table 8, resorting to the relation (4), one can obtain masses of the spin-2, spin-4, spin-6, and spin-8 states lying on these trajectories. The results extracted by the square-root trajectory are shown in Table 9.

TABLE 9: Comparison of the masses of the $J = 2, 4, 6, 8$ mesons lying on the $1^3P_2$ trajectories with the experimental data. (All in MeV.)

|  | $J = 2$ | | | $J = 4$ | | | $J = 6$ | | | $J = 8$ | | |
| --- | --- | --- | --- | --- | --- | --- | --- | --- | --- | --- | --- | --- |
|  | Quasi-linear | Square-root | Exp.[21] | Quasi-linear | Square-root | Exp.[21] | Quasi-linear | Square-root | Exp.[21] | Quasi-linear | Square-root | Exp.[21] |
| $M_{n\bar{n}}$ | **1318.3** | **1318.3** | 1318.3 | **1995** | **1995** | 1995 | 2494 | 2353 | 2450 | 2909 | 2531 |  |
| $M_{n\bar{s}}$ | **1429.0** | **1429.0** | 1429.0 | **2045** | **2045** | 2045 | 2514 | 2374 |  | 2909 | 2531 |  |
| $M_{c\bar{n}}$ | **2463.5** | **2463.5** | 2463.5 | 3060 | 3055 |  | 3558 | 3451 |  | 3994 | 3715 |  |
| $M_{c\bar{s}}$ | 2540 | 2547 | 2571.9 | 3106 | 3106 |  | 3583 | 3482 |  | 4004 | 3730 |  |
| $M_{c\bar{c}}$ | **3556.20** | **3556.20** | 3556.20 | 4122 | 4115 |  | 4620 | 4531 |  | 5069 | 4842 |  |
| $M_{b\bar{n}}$ | **5738** | **5738** | 5738 | 6204 | 6155 |  | 6636 | 6493 |  | 7043 | 6764 |  |
| $M_{b\bar{s}}$ | 5771 | 5788 | 5839.83 | 6226 | 6194 |  | 6649 | 6522 |  | 7048 | 6784 |  |
| $M_{b\bar{c}}$ | 6799 | 6798 |  | 7269 | 7225 |  | 7710 | 7583 |  | 8128 | 7881 |  |
| $M_{b\bar{b}}$ | **9912.21** | **9912.21** | 9912.21 | 10351 | 10293 |  | 10771 | 10627 |  | 11176 | 10920 |  |

## 5. Discussions and Summary

In this work, in the framework of quasi-linear Regge trajectory and square-root Regge trajectory, the parameters of the $1^1S_0$, $1^3S_1$, and $1^3P_2$ trajectories are extracted. Based on these parameters, the masses of the orbital excited states lying on these trajectories mentioned above are estimated and shown in Table 3, Table 6, and Table 9, respectively. In the analysis of Ref. [5], the masses of mesons were estimated by the quasi-linear Regge trajectory with the assumption that the slopes of the parity partners trajectories coincide. In the analysis of Ref. [20], the masses of mesons were calculated by the square-root Regge trajectory with the assumption that the threshold parameter of the parity partners trajectories coincide. In our consideration, we do not adopt the above assumption, but resort to the experimental data.

In Table 3, we can see that the masses of the ground states $c\bar{s}$, $b\bar{s}$, and $b\bar{c}$ extracted by the quasi-linear trajectory and square-root trajectory are almost equal. Furthermore, based on the assignment from PDG, $D_s^*$ and $B_s^*$ are assigned to the candidates of the ground states $c\bar{s}$ and $b\bar{s}$. Comparison of their masses given by the two trajectories with the experimental data illustrates that there is a good agreement between the results of the two trajectories and the experimental data. In the 2016 updated Meson Summary Table [21], $D_{s3}^*(2860)$ is assigned to the candidate of $c\bar{s}$ of $1^3D_3$ meson multiplet, i.e. the candidate of $c\bar{s}$ of J=3. The predictions given by the quasi-linear trajectory and square-root trajectory for the mass of the above candidate are 2815 MeV and 2807 MeV, which is consistent with the experimental value 2863.2 MeV. According to PDG, the orbital excited states $\rho_5(2350)$ and $K_5^*(2380)$ are assigned to the candidates of $n\bar{n}$ and $n\bar{s}$ of J=5. The masses given by the quasi-linear trajectory and square-root trajectory for the state $n\bar{n}$ of J=5 are 2259 MeV and 2102 MeV, respectively. Comparison with the experimental value 2330 MeV implies that the result extracted by the quasi-linear trajectory is of better agreement. The assignment of $K_5^*(2380)$ (2382 MeV), strongly suggest that the result of the quasi-linear trajectory (2347 MeV) is in better agreement than the square-root trajectory (2196 MeV). $X(2750)$ was observed at SLAC with mass $M = 2747 \pm 32$ MeV, isospin I = 1, spin-parity $J^P = 7^-$, and suggested to be the isovector member of $1^3I_7$ meson multiplet [33]. This assignment was supported in Refs. [34-36]. In the presence of this assignment of high-spin state $X(2750)$, the result estimated by the quasi-linear trajectory (2712 MeV) is of better agreement than the result of the square-root trajectory (2301 MeV).

In Table 6, we can see that the results given by the two trajectories for the ground states $c\bar{s}$, $b\bar{s}$ and $b\bar{c}$ are almost equal and accord well with the experimental data. In Table 9, for the ground states $c\bar{s}$, $b\bar{s}$ and $b\bar{c}$, consistently suggest that the results of the two trajectories are very close and accord well with the experimental data. Moreover, in Table 9, the orbital excited state $a_6(2450)$ is assigned to the candidate of $n\bar{n}$ of J=6 [21]. The masses given by the quasi-linear trajectory and square-root trajectory for the candidate are 2494 MeV and 2353 MeV, respectively. Obviously the result given by the quasi-linear trajectory is more close to the experimental value 2450 MeV.

In addition, the square-root trajectory has a limit for higher spins, i.e. $J_{max} = [J(T_{i\bar{j}})]$ [20], while the quasi-linear Regge trajectory does not have the above limit. According to the relation (4), we calculate the $J_{max}$ of the $1^1S_0$, $1^3S_1$, and $1^3P_2$ trajectories, the results are shown in Table 10-12. So the discovery of the high spin states $J > J_{max}$ on experiment, would test whether the square-root trajectory is right or not.

TABLE 10: $J_{\max}$ of $1^3S_1$ square-root Regge trajectories.

| | $n\bar{n}$ | $n\bar{s}$ | $c\bar{n}$ | $c\bar{s}$ | $c\bar{c}$ | $n\bar{b}$ | $s\bar{b}$ | $c\bar{b}$ | $b\bar{b}$ |
|---|---|---|---|---|---|---|---|---|---|
| $J_{\max}$ | 8 | 8 | 11 | 12 | 15 | 17 | 18 | 20 | 26 |

TABLE 11: $J_{\max}$ of $1^1S_0$ square-root Regge trajectories.

| | $n\bar{n}$ | $n\bar{s}$ | $c\bar{n}$ | $c\bar{s}$ | $c\bar{c}$ | $n\bar{b}$ | $s\bar{b}$ | $c\bar{b}$ | $b\bar{b}$ |
|---|---|---|---|---|---|---|---|---|---|
| $J_{\max}$ | 9 | 9 | 11 | 12 | 14 | 18 | 18 | 21 | 27 |

TABLE 12: $J_{\max}$ of $1^3P_2$ square-root Regge trajectories.

| | $n\bar{n}$ | $n\bar{s}$ | $c\bar{n}$ | $c\bar{s}$ | $c\bar{c}$ | $n\bar{b}$ | $s\bar{b}$ | $c\bar{b}$ | $b\bar{b}$ |
|---|---|---|---|---|---|---|---|---|---|
| $J_{\max}$ | 9 | 9 | 12 | 12 | 15 | 17 | 17 | 20 | 25 |

Based on the above discussions, for the ground states, the results given by the quasi-linear trajectory and square-root trajectory are very close, which accord well with the experimental data. For the orbital excited states, with the $J$ quantum number increases, the meson masses given by the quasi-linear trajectory are higher than the square-root trajectory. Comparison of the masses of orbital excited states $\rho_5(2350)$, $K_5^*(2380)$, $a_6(2450)$, and $X(2750)$ with those given by the quasi-linear trajectory and square-root trajectory implies that the quasi-linear trajectory describe the existing meson spectrum to be more reasonable. Searching for highly orbital excited mesons (especially for $J > J_{\max}$ mesons) is very important to identify whether Regge trajectory is quasi-linear or square-root form. We suggest more efforts should be given to search highly orbital excited mesons experimentally.

## Acknowledgments

We are grateful to Yao-Feng Zhang (Beijing Normal U.) for valuable discussions. This work was supported in part by the National Natural Science Foundation of China (Grant no. U1204115 and no. 11605009), the Yili Normal University Foundation of China (Grant no. 2016YSYB08), the Xinjiang Natural Science Foundation of China (Grant no. 2016D01C384).